\documentclass[aps,prl,a4paper,superscriptaddress,12pt,preprint,floatfix,footinbib]{revtex4-1}
\usepackage{amsmath}
\usepackage{amsfonts}
\usepackage{amssymb}
\usepackage{graphicx}
\usepackage{epstopdf}
\usepackage[colorlinks=true,linkcolor=blue,citecolor=blue,urlcolor=blue]{hyperref}
\usepackage{setspace}
\usepackage{multirow}
\usepackage{physics}

\begin{document}

\title{Path-dependent initialization of a single quantum dot exciton spin in a nano-photonic waveguide}
\author{R.J. Coles}
\email[]{r.j.coles@sheffield.ac.uk}
\author{D.M. Price}
\author{B. Royall}
\affiliation{Department of Physics and Astronomy, University of Sheffield, Sheffield, S3 7RH, UK.}
\author{E. Clarke}
\affiliation{EPSRC National Centre for III-V Technologies, Department of Electronic and Electrical Engineering, University of Sheffield, Sheffield S1 3JD, UK}
\author{M.S. Skolnick}
\author{A.M. Fox}
\author{M.N. Makhonin}
\email[]{m.makhonin@sheffield.ac.uk}
\affiliation{Department of Physics and Astronomy, University of Sheffield, Sheffield, S3 7RH, UK.}

\begin{abstract}
\noindent 
 We demonstrate a scheme for in-plane initialization and readout of a single exciton spin in an InGaAs quantum dot (QD) coupled to a GaAs nanobeam waveguide. The chiral coupling of the QD and the optical mode of the nanobeam enables spin initialization and readout fidelity approaching unity in magnetic field $B=1$ T and $\sim$0.9 without the field. We further show that this in-plane excitation scheme is independent of the incident excitation laser polarization and depends solely on the excitation direction. This scheme provides a robust in-plane spin excitation basis for a photon-mediated spin network for quantum information applications.
\end{abstract}


\maketitle

The spin states of quantum dots (QDs) have been recognised as promising solid-state qubits for applications in quantum information processing,
with a number of key operations having already being demonstrated using off-chip excitation schemes \cite{Warburton2013,Delteil2015}. However, the development of scalable and compact spin networks \cite{Ramos2016,Vermersch2016} based on QDs requires that the dots should be embedded on-chip within photonic structures. This has the added benefit of enhancing the light-matter interaction and providing an efficient mechanism for the generation and manipulation of single photons on-chip \cite{Lodahl2013b}. The establishment of spin networks then relies on demonstrating faithful communication of spin information between matter qubits (QD spins) via flying qubits (photons) within a photonic circuit \cite{DiVincenzo2000,Barrett2004}. To this end, an efficient on-chip spin-photon interface is required to map the spin of the matter qubit onto the spin or direction of the flying qubit.

Directional emission is an example of optical spin-orbit coupling, and is a general property of wavelength-scale optical systems due to the existence of elliptically polarized electric fields in guided modes \cite{Bliokh2015,Lodahl2016}.
Such uni-directional phenomena have been observed using a variety of physical systems using several different types of nano-photonic structures \cite{Coles2016,Junge2013,LeFeber2015,Luxmoore2013,Luxmoore2013a,Mitsch2014a,Petersen2014,Rodriguez-Fortuno2014,Rodriguez-fortuno2013,Sollner2015a}. 
These experiments demonstrate spin-to-path conversion where the local polarization or photon spin is directly mapped to the propagation direction in the photonic structure. The reciprocal effect, namely photon path to spin conversion, has been demonstrated in a passive silicon photonic device consisting of a microdisk coupled to a waveguide \cite{Rodriguez-Fortuno2014} and for a single $^{85}$Rb atom evanescently coupled to a whispering gallery mode resonator \cite{Junge2013}. However, path-to-spin based initialization of a solid-state quantum emitter remains to be demonstrated.

In this Letter we demonstrate such a scheme for QD exciton spin initialization and readout in a single-mode nanobeam waveguide exploiting the reciprocal (path-to-spin) coupling effect. In previous work we demonstrated spin-to-path conversion for a QD coupled to the circularly polarized fields of a nanobeam by observing uni-directional spin-dependent single photon emission \cite{Coles2016}. Here we show that a resonant photon propagating in a specific direction excites a correspondingly polarized exciton spin state when absorbed by a QD. This direction-dependent initialization of a single QD exciton spin is a key goal towards the realization of an on-chip quantum spin network and is the focus of this paper.

\begin{figure*}
\includegraphics[width=\textwidth]{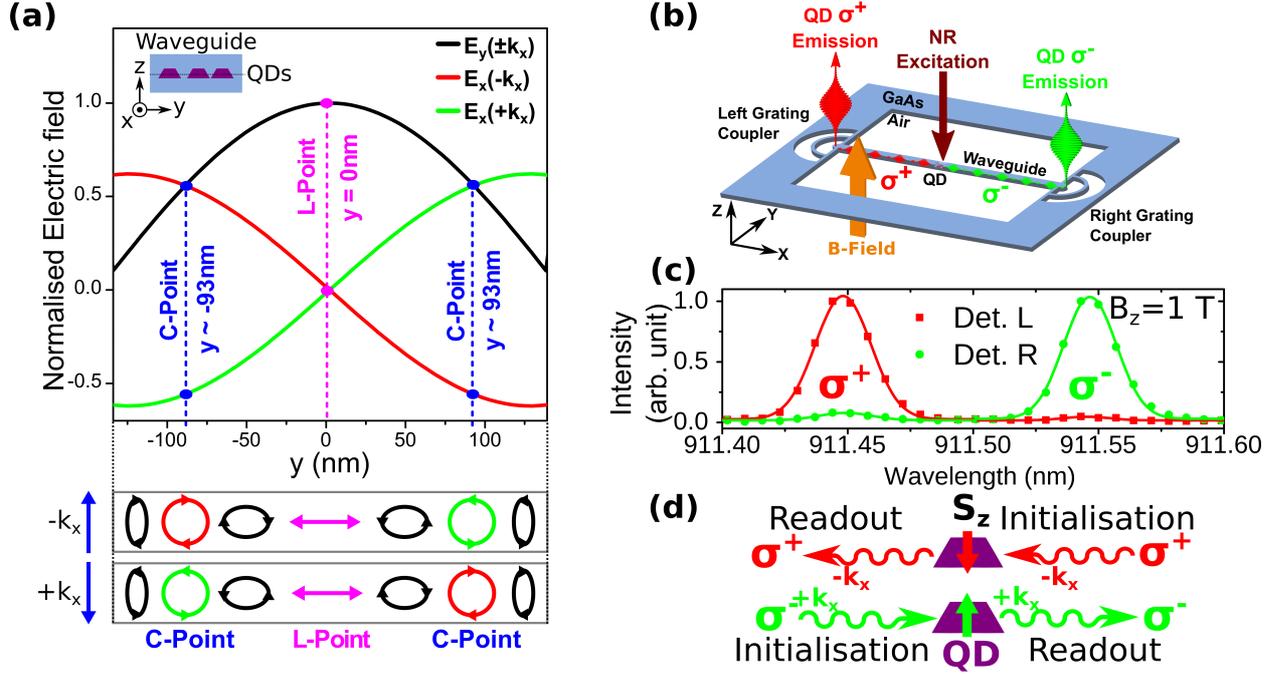}
\caption{\label{fig1}(color online). (a) \textit{Upper:} Internal electric field profiles, $E_x$ (red and green) and $E_y$ (black), of the waveguide mode propagating along the x-axis in the QD plane at $z=0$. A relative phase of $|\phi|=\pi/2$ exists between the two components. (inset) Cross section of the waveguide. \textit{Lower:} Helicity of the modal fields with lateral displacement from the waveguide center, $y$, for each propagation direction. (b) Schematic of the device structure and experimental arrangement for QD exciton spin readout. (c) PL spectra of the chirally coupled QD taken at $B_z=1$ T when detecting emission from each of the two grating couplers (left,right labelled and shown with corresponding color). (d) Schematic illustrating the relationship between photon polarization ($\sigma^{+},\sigma^{-}$), propagation direction ($-k_x,+k_x$), and exciton spin ($\uparrow,\downarrow$) for optical readout and initialization of the exciton spin.}
\end{figure*}

Our scheme for QD exciton spin readout and initialization is presented in Fig. \ref{fig1}, which is based on the chiral coupling of the QD to the waveguide mode \cite{Coles2016}. Analysis of the modal fields reveal elliptical in-plane electric fields, where the helicity depends on the propagation direction [Fig. \ref{fig1}(a)]. If a QD is located or deterministically placed at a chiral point, where the electric field is circular (denoted C-point), the exciton spins will couple to photons which propagate in opposite directions with circular polarization transverse to the propagation plane  with unity fidelity \cite{Coles2016}.

Our waveguide device consists of a vacuum-clad single mode GaAs nanobeam waveguide containing InGaAs QDs in the xy-plane at z=0 [see inset of Fig.\ref{fig1}(a)]. Grating couplers \cite{Faraon2008} are added to each end of the nanobeam [Fig.\ref{fig1}(b)] to allow free space coupling to the waveguide mode \cite{Coles2016,Arcari2014,Makhonin2014,Prtljaga2014, Sollner2015a,Note1}. The waveguide possesses quasi-continuous translational symmetry so the constraints on the emitter location along $x$ are relaxed relative to devices based upon photonic crystal waveguides \cite{Sollner2015a,Young2015}. This allows for a higher probability of chirally-coupled QDs when randomly positioned and a larger tolerance in the position for deterministically positioned QDs, as shown in \cite{Coles2016}. The sample is held at 4 K in a liquid helium bath cryostat and a magnetic field $B_z$ can be applied normal to the device plane. A confocal microscopy system with motorized scanning mirrors allows independently spatially resolved excitation and detection of photoluminescence (PL) in the sample. For further details on the experimental setup see the supplemental material in \cite{Note1}.

Selection of a QD, from a randomly positioned ensemble, which exhibits uni-directional emission was first conducted using PL spectroscopy with non-resonant (NR) laser diode excitation at 808 nm under an applied magnetic field of $B_z=1$ T, as shown schematically in Fig. \ref{fig1}(b). The magnetic field lifts the degeneracy of the spin eigenstates of the QD exciton through the Zeeman effect and facilitates their identification via emission of circularly polarized photons of different wavelength. \mbox{Figure \ref{fig1}(c)} shows PL spectra of a chirally-coupled QD when detecting PL emission from the grating couplers. Strong emission from respective Zeeman components ($\sigma^{+}/\sigma^{-}$) is observed from each grating (left/right) with Zeeman energy splitting \mbox{$\Delta E_z = 147$ $\mu$eV}. The spin readout contrast is derived as in \cite{Coles2016} from the relative intensity of the Zeeman components $I^{\sigma +}$ and $I^{\sigma -}$ using

\begin{equation}\label{readout_contrast_eq}
C^{read}_{det. l/r} = \frac{I^{\sigma +}_{l/r} - I^{\sigma -}_{l/r}}{I^{\sigma +}_{l/r} + I^{\sigma -}_{l/r}},
\end{equation}

\noindent where the subscripts l/r refer to the left and right coupler from which PL emission is detected. Readout contrasts of $C_{det. l}^{read}=0.95\pm0.05$ and $C_{det. r}^{read}=-0.88\pm0.06$ are observed from the left and right couplers respectively. The small discrepancy in contrasts between left and right detection may arise from a slight asymmetry in the reflectivity of the grating couplers. Hanbury-Brown Twiss (HBT) measurements, presented in \cite{Note1}, confirm that the PL emission in Fig. \ref{fig1}(c) comprise antibunched single photons.

The observation of high contrast uni-directional emission in Fig. \ref{fig1}(c) shows that the spin states of the exciton are well coupled to waveguide modes which propagate in opposite directions. We conclude that the QD resides very close to a C-point and exhibits spin to path conversion. This process is illustrated schematically in \mbox{Fig. \ref{fig1}(d)}. As described in more detail in the supplemental material in \cite{Note1}, the QD transition studied corresponds to $X^+$ trion recombination. The spin states of $X^{+}$ are $\ket{\downarrow\Uparrow\Downarrow}$ and $\ket{\uparrow\Uparrow\Downarrow}$, where $\uparrow,\downarrow$ and $\Uparrow,\Downarrow$ denote electron and heavy hole spins respectively. These states correspond to residual hole spin states $\ket{\Downarrow}$ and $\ket{\Uparrow}$ where $\sigma^{+}$ and $\sigma^{-}$ photons are emitted, on exciton recombination, to the left and right waveguide directions correspondingly. The use of a charged exciton is ideal for a spin-photon interface as selection rules dictate that the polarization of emitted photons is always circular which therefore directly maps the exciton spin to the helicity of photon polarization. In addition, the $X^+ \rightarrow h^+$ transition allow for coherent initialization and control of the residual hole spin \cite{Kroner2008,Godden2012,DeGreve2011,Brash2015}. 

\begin{figure}
\includegraphics[width=\columnwidth]{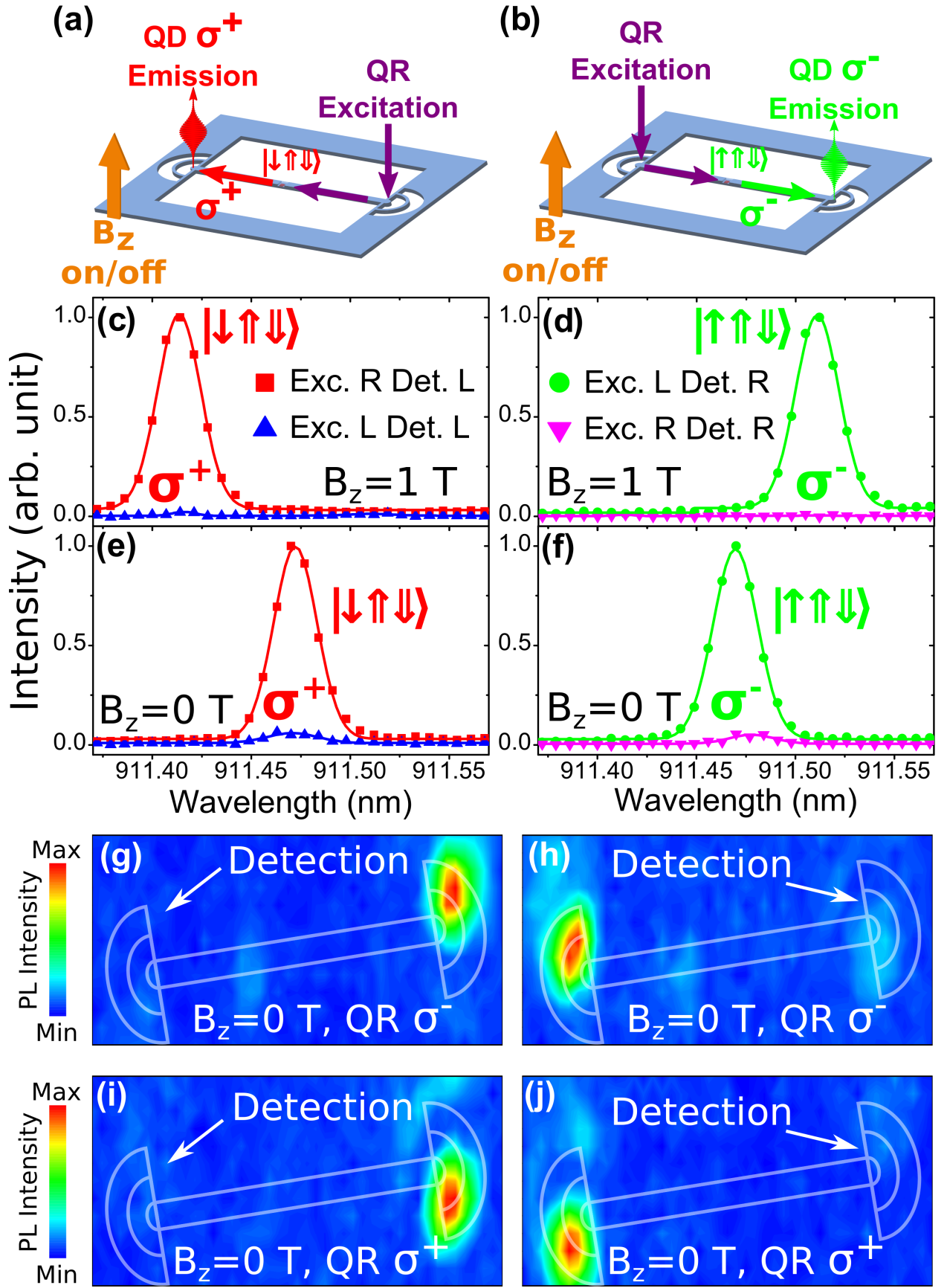}
\caption{\label{fig2}(color online). (a)-(b) Schematic of the experimental arrangement for QD exciton spin initialization. The QD is at the chiral point in the waveguide. (c)-(d) PL spectra obtained at $B_z=1$ T when quasi-resonantly exciting the QD from each grating coupler with detection fixed at the (c) left and (d) right coupler. (e)-(f) PL spectra as in (c),(d) but with $B_z=0$ T. The legend refers to the couplers on which the QR excitation laser and PL detection were located respectively (L=left coupler, R=right coupler). Spectral data are shown as discrete points, with Gaussian fits as solid lines. (g)-(h) Excitation maps obtained by performing a raster scan of the $\sigma^-$ polarized QR laser over the device whilst keeping detection fixed at the (g) left and (h) right coupler. (i)-(j) Excitation maps as in (e),(f) but with $\sigma^+$ laser polarization. All maps recorded PL emission filtered at the QD wavelength and a schematic outline of the waveguide and couplers are overlaid. The apparent rotation of the device from horizontal is due to a small rotation induced by the detection optics.}
\end{figure}

By reciprocity, it is expected that a resonant photon propagating in a given direction in the waveguide will excite only one of the two exciton spin states. However, distinction between photons emitted by the QD and from a resonant laser in a waveguide geometry is nontrivial \cite{Makhonin2014}. We note that quasi-resonant (QR) p-shell excitation of the QD exciton has been shown to preserve the spin of the incident photon and to initialize the exciton spin eigenstates with fidelity of $\sim0.95$ \cite{DelaGiroday2010}, whilst allowing for effective spectral filtering of QD PL emission. Therefore we employ a single frequency diode laser coupled to the p-shell resonance of the $X^+$ trion via the grating coupler (identification of the p-shell resonance is presented in the supplemental material in \cite{Note1}). QR laser photons are scattered from the coupler along the waveguide towards the QD at a C-point, where the photon spin is transferred to the spin state of the p-shell $X^+$ trion on absorption. An LO phonon assisted process allows the carriers to relax rapidly to the ground state in a spin conserving process. Recombination of the s-shell $X^+$ trion then produces a $\sigma^+/\sigma^-$ polarized photon which propagates either to the left or right coupler depending on the initial spin state. A schematic illustrating this process is presented in Fig. \ref{fig1}(d) and experimental arrangements are presented in \mbox{Fig. \ref{fig2}(a)-(b)}. PL emission from the QD is only observed under QR excitation when weak NR excitation is also applied which populates and stabilises the residual charge within the QD (e.g. as in \cite{Makhonin2014}). The PL contribution from the NR laser is weak and constant in all spectra, so it is subtracted from the data as discussed in the supplemental material in \cite{Note1}.

Figure \ref{fig2}(c) presents exciton PL emission spectra measured from the left coupler for QR excitation on either coupler with $B_z=1$ T. When QR excitation is applied to the right coupler to a good approximation only the $\sigma^{+}$ emission peak is seen. Conversely, for detection on the right coupler, the $\sigma^{-}$ peak is observed only when QR excitation is applied to the left coupler, as shown in Fig. \ref{fig2}(d). In both cases only very weak signal is observed with excitation and detection on the same coupler. These results demonstrate the initialization of the exciton spin using path-spin conversion and subsequent directional re-emission, which can be understood as follows. When QR excitation is applied to the right coupler, the waveguide mode propagating from right to left is excited. Weak NR excitation of the QD located at a C-point creates a resident hole with random spin. If this hole is in the $\ket{\Downarrow}$ state, the QD absorbs a $\sigma^+$ polarized QR photon from the left propagating waveguide mode which creates an $X^+$ trion with spin state $\ket{\downarrow\Uparrow\Downarrow}$. This exciton then recombines and emits a $\sigma^+$ photon into the same waveguide direction leaving a residual hole in state $\ket{\Downarrow}$. The photon is detected from the left coupler as illustrated in Fig. \ref{fig2}(a). If the hole is initially in the $\ket{\Uparrow}$ state, the photon is not absorbed by the QD. When QR excitation is moved to the left coupler, the exciton is prepared in the $\ket{\uparrow\Uparrow\Downarrow}$ spin state and the emitted $\sigma^-$ photon propagates to the right coupler, as illustrated in Fig. \ref{fig2}(b), leaving a residual hole state $\ket{\Uparrow}$. In the absence of exciton spin relaxation within the QD, emitted photons from the same coupler where the excitation is applied are not expected, exactly as observed. Therefore the device exhibits path-spin and spin-path conversion $\ket{L} \leftrightarrow \ket{\downarrow\Uparrow\Downarrow}$ and $\ket{R} \leftrightarrow \ket{\uparrow\Uparrow\Downarrow}$ where L and R are the propagation directions of the photon.

To quantify the fidelity of spin initialization, we define the initialization contrast for detection on each grating coupler as

\begin{equation}\label{init_contrast_eq}
C^{init}_{det. l/r} = \frac{(I_{R/L}^{\sigma^+}+I_{R/L}^{\sigma^-})-(I_{L/R}^{\sigma^+}+I_{L/R}^{\sigma^-})}{(I_{R/L}^{\sigma^+}+I_{R/L}^{\sigma^-})+(I_{L/R}^{\sigma^+}+I_{L/R}^{\sigma^-})},
\end{equation}

\noindent where the superscripts $\sigma^{+/-}$ refer to the peaks in \mbox{Fig. \ref{fig2}(c)-(f)}, the upper case subscripts $L/R$ refer to excitation on either the left or right coupler and the lower case subscripts $l/r$ refer to detection on either the left or right coupler. For this QD we extract $C^{init.}_{det. l}=0.96\pm0.05$ and $C^{init.}_{det. r}=-0.99\pm0.06$. The small difference in initialization contrasts is attributed to the intrinsic asymmetry of the QD emission as observed in Fig. \ref{fig1}(c). HBT measurements are presented in \cite{Note1} for the PL emission lines in Fig. \ref{fig2}(c)-(f) which confirm that the single photon emission character is retained under QR excitation.

Near perfect initialization and readout contrasts highlight the practicality of the device as a spin-photon interface. When one considers the application of the device, however, the requirement for an external magnetic field may present a barrier to scalability under some circumstances. We therefore repeated the experiments for $B_z=0$ T and the resulting spectra are presented in \mbox{Figs. \ref{fig2}(e) and \ref{fig2}(f)}. Large initialization contrasts in the absence of magnetic field were observed with $C^{init.}_{det. l}=0.88\pm0.03$ and $C^{init.}_{det. r}=-0.91\pm0.03$. The $\sim8\%$ lower contrast values are attributed to the greater contribution of the spin hyperfine interaction with nuclei comprising the QD \cite{Chekhovich2013} at $B_z=0$ T.

To demonstrate that the initialised spin is determined by the direction of excitation only and not the incident QR laser polarization we provide excitation maps for different QR laser polarizations. The maps are obtained using a raster scan of QR excitation over the device whilst keeping the detection fixed at one grating coupler, filtered at the QD emission wavelength. The results of these measurements are presented in Figs. \ref{fig2}(g) and \ref{fig2}(h) for detection on the left and right couplers respectively using $\sigma^-$ polarized excitation, whilst in Figs. \ref{fig2}(i) and \ref{fig2}(j) a $\sigma^+$ polarized laser is used. As can be clearly seen from these data, intense PL emission from the QD is observed only when exciting and detecting from opposite couplers; the initialization does not depend on the laser polarization. Since the chiral coupling of the QD and the waveguide mode is determined by the local electric field polarization of the waveguide mode, the initialization contrast is expected to be independent of the QR polarization. The polarization of the incident laser only determines the overlap with the free space optical modes of the grating coupler and hence the efficiency with which the waveguide mode is excited. The independence of the waveguide mode on the incident laser polarization is discussed in more detail in the supplemental material in \cite{Note1}. 

\begin{figure}
\includegraphics[width=\columnwidth]{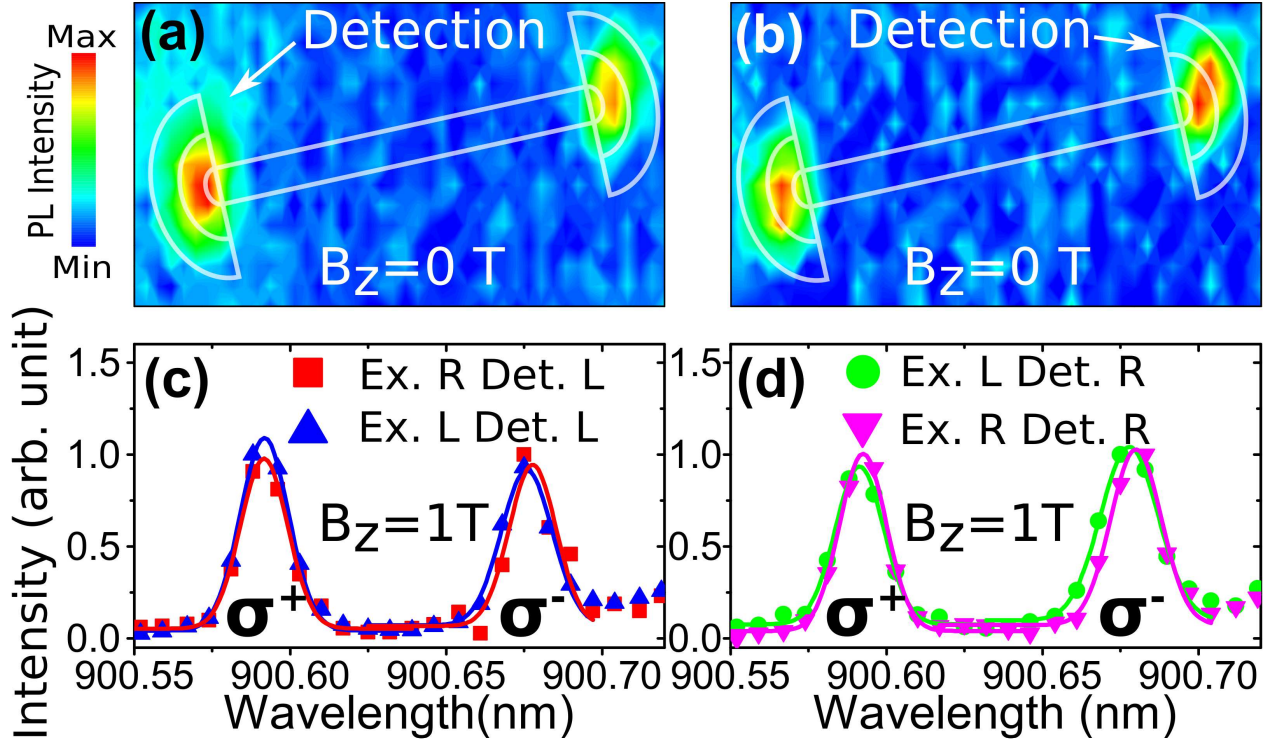}
\caption{\label{fig3}(color online). (a)-(b) Excitation maps obtained by a raster scan of QR excitation over a device containing a non-chirally coupled QD whilst keeping detection fixed at the (a) left and (b) right grating couplers at $B_z=0$ T. All maps recorded PL emission filtered at the QD wavelength and a schematic outline of the waveguide and couplers are overlaid. The apparent rotation of the device from horizontal is due to a small rotation induced by the detection optics. (c)-(d) Corresponding PL spectra obtained when exciting the QD from each coupler with detection fixed at the (c) left and (d) right coupler when an external magnetic field of $B_z=1$ T is applied. Spectral data are shown as discrete points, with an experimental Gaussian fit to the data shown as a solid line.}
\end{figure}

To verify that the high degree of spin initialization arises from the chiral coupling of the QD with the waveguide mode, measurements were also made on a QD which does not exhibit uni-directional emission. For this to occur the QD must be located close to the centre of the waveguide, where the modal fields are linear [L-points, see Fig. \ref{fig1}(a)]. The QD that is symmetrically coupled is identified from NR excitation measurements first as in Fig.1(b) with the main difference that both $\sigma^{+}/\sigma^{-}$ are detected with equal intensity from both couplers. The results of the QR excitation experiments on this non-chiral QD are presented in Fig. \ref{fig3}. The excitation maps in Fig. \ref{fig3}(a)-(b) show almost equal emission intensities when the QD is excited from either grating coupler, independent of which coupler the emission is detected from. These findings are reflected in the measured spectra of Fig. \ref{fig3}(c)-(d), where extracted initialization contrasts very close to zero are found with $C^{init.}_{det. l}=(9\pm1)\times10^{-3}$ and $C^{init.}_{det. r}=(-13\pm1)\times10^{-3}$ These results confirm that the high fidelity of spin-path conversion for the displaced QD is indeed due to chiral coupling between the exciton emission and the waveguide mode.

To conclude, we demonstrate in-plane initialization and readout of a charged exciton spin in a QD chirally coupled (located at a C-point) to a single mode waveguide using a quasi-resonant excitation scheme. Spin initialization is demonstrated with high directionality contrasts $\sim1$ at $B_z=1$ T and $\sim0.9$ at $B_z=0$ T. We compare with a non-chiral QD (located at L-point) and show that the high fidelity of spin preparation is indeed due to the chiral coupling of the QD. Furthermore, we show that the addressing of each exciton spin state is determined by the excitation path and is independent of the incident laser polarization. The findings and techniques presented establish a method for communication between two or more quantum dots on-chip and have potential to contribute to the creation of spin-optical on-chip networks. 

\begin{acknowledgments}
The authors gratefully acknowledge funding from \mbox{EPSRC} under research grant number EP/J007544/1.
\end{acknowledgments}


%

\end{document}